\documentclass{ptephy}

\usepackage{amsmath}
\usepackage{graphicx}
\usepackage{subfigure}

\newcommand{\ket}[1]{\lvert #1 \rangle}
\newcommand{\braket}[1]{\langle#1 \rangle}

\begin{document}

\title{Bulk Property on Cayley Tree with Smooth Boundary Condition}
\author{
\name{\fname{Hiro-Aki} \surname{Hotta}}{}
}
\address{
Department of Physics, Graduate School of Science, Kobe University, Kobe 657-8501, Japan
\email{hiroaki.hotta@stu.kobe-u.ac.jp}
}
\subjectindex{A21, A64}

\begin{abstract}
We study a nearest-neighbor hopping model on the Cayley tree under the smooth boundary condition with the modulation function $f_s=\sin^2[\pi s/(2M+1)]$, where $s$ is a distance from the central site, and $M$ is the number of shells on the tree.
As a result of this smoothing, the particle density in the ground state becomes nearly uniform in the bulk region even when $M$ is relatively small.
We compare the calculated particle density at the center with exact result on the Bethe lattice, and they show a good agreement.
The calculated bond energy at the center also agrees with that on the Bethe lattice.
\end{abstract}

\maketitle

\section{Introduction}
The Cayley tree is a loopless graph where the coordination number $z$ on each site is the same except at the boundary.
It is known that boundary effects tend to be dominant on the Cayley tree, since the number of the boundary sites on the tree increases exponentially with respect to the diameter $M$ of the tree.
In some cases, the boundary effect remains even in the thermodynamic limit $M\rightarrow \infty$, as is known in the case of ferromagnetic Ising model\cite{Runnels, Eggarter, Matsuda, Morita, Muller-Hartmann, Baxter};
a thermodynamic quantity of the whole system is dominated by those area near the boundary.
On the other hand, an expectation value of a local quantity at the center of the tree can be different from the averaged value for the entire system. 
Such a ``bulk" property, which appears deeply inside the system, can be described by the Bethe lattice, which can be regarded as an infinite Cayley tree without boundary.

In this article, we consider a way of obtaining the ``bulk" property of observables on the Cayley tree.
For this purpose we try to suppress the boundary effect.
It is well known that the strength of the boundary effects is dependent on boundary conditions.
The most simple one is the open boundary condition (OBC), where the boundary effect is relatively conspicuous.
A typical example is the Friedel oscillation caused by the open boundary.
It is possible to suppress such a sudden boundary effect by imposing the periodic boundary condition (PBC).
However, PBC on the Cayley tree is non-trivial.
A solution is to introduce so called the smooth boundary condition (SBC)\cite{sbc1, sbc2}, where the local energy scale varies smoothly from the maximum at the center to zero at the boundary.

In one dimension, a class of SBC called ``sine-square deformation" (SSD)\cite{ssd} efficiently reduces the boundary effect.
For a variety of one-dimensional systems, it has been confirmed that the boundary effect in the ground state vanishes entirely under SSD\cite{ssd,ssd-HN, ssd-K1, ssd-GDLN, ssd-K2, ssd-SH}.
The SSD has been applied to quantum entanglement analysis on spin chains\cite{Hikihara} and the ground state study of a simple string theory\cite{Tada}.
The validity of SSD for two-dimensional system is also confirmed\cite{ssd-MKH, Hotta1, Nishimoto}.
Recently, a method of obtaining bulk quantities by means of SSD, the method which is called as the grand canonical analysis, is proposed\cite{Hotta1,Hotta2} and applied to kagome Heisenberg antiferromagnet\cite{Nishimoto}.
An advantage of this grand canonical method is that one can easily access bulk quantities from numerical observables of finite-size systems.

In this paper, we compare bulk properties of the nearest-neighbor hopping model on the Bethe lattice, with observables on the Cayley tree, performing a numerical calculation with the aid of the grand canonical analysis.
This paper is organized as follows.
In Sec.~\ref{sec:TB}, we introduce the nearest-neighbor hopping model on the Cayley tree.
We shall see that the Friedel oscillation disturbs the bulk quantity under OBC.
In Sec.~\ref{sec:SBC}, we impose SBC to the model.
The obtained results are compared with the bulk quantities on the Bethe lattice.
We conclude the obtained result in Sec.~\ref{sec:last}.

\section{Nearest-Neighbor Hopping Model on Cayley Tree} \label{sec:TB}
\begin{figure}[htb]
\centering
\includegraphics[width=0.4\textwidth,clip]{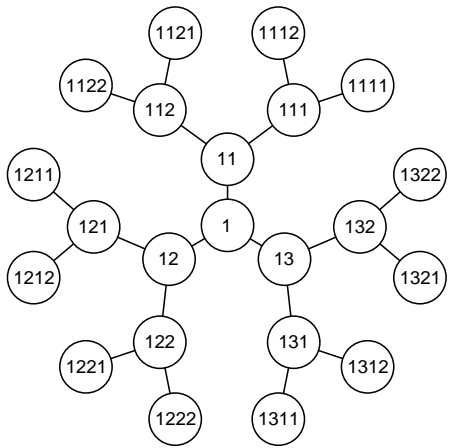}
\caption{A Cayley tree with coordination number with $z=3$. The number of shell $M$ is equal to 3. All sites are labeled according to the index rule in Eq.~\eqref{labeling}.}
\label{tree}
\end{figure}

We first consider how to label sites on the Cayley tree.
We introduce a site-centered tree with coordination number $z$, where there are $M$ shells.
Figure~\ref{tree} shows the case $z=3$ and $M=3$.
All the interior sites have $z$ neighboring sites, but those boundary sites have only one.
Regarding the center of the lattice as the 0-th shell, each site on the $s$-th shell can be labelled by a set of $s+1$ integers
\begin{align}
R=\{ r_0\: r_1\ldots r_s \},
\label{labeling}
\end{align}
where $s=0,1,\ldots,M$, and where $r_i$ takes values as follows
\begin{align}
r_i =
\begin{cases}
1 & (i=0)\\
1,\ldots, z & (i=1)\\
1,\ldots, z-1 & (i=2,\ldots,M)
\end{cases}
.
\end{align}

Next we introduce a spinless nearest-neighbor hopping model on the Cayley tree.
We introduce fermionic creation and annihilation operator, respectively, $c^{\dag}_{R}$ and $c^{}_{R}$ on each site labeled by $R$ in Eq.~\eqref{labeling}.
The hopping between neighboring sites $R$ and $R'$ is expressed as
\begin{align}
h(R, R') = -t (c^{\dag}_{R} c^{}_{R'} + c^{\dag}_{R'} c^{}_{R}),
\end{align}
where $t$ corresponds to the hopping amplitude, and the on-site potential is given by
\begin{align}
g(R) = -\mu c_{R}^{\dag} c^{}_{R},
\end{align}
where $\mu$ denotes the chemical potential.
By means of these notations, the nearest-neighbor hopping Hamiltonian on the tree is given by
\begin{align}
H =\sum_{s=0}^{M-1} \sum_{r_0\ldots r_{s+1}} h(\{ r_0\ldots r_s \}, \{ r_0\ldots r_{s+1} \}) +\sum_{s=0}^{M} \sum_{r_0\ldots r_s} g(\{ r_0\ldots r_s \}).
\label{H_OBC}
\end{align}
For example, in the case of $z=3,M=2$, the Hamiltonian is written as
\begin{align}
H = & h(\{1\}, \{11\}) + h(\{1\}, \{12\}) + h(\{1\}, \{13\}) + g(\{1\}) \notag\\
& + h(\{11\}, \{111\}) + h(\{11\}, \{112\})  + g(\{11\}) \notag\\
& + h(\{12\}, \{121\}) + h(\{12\}, \{122\}) + g(\{12\}) \notag\\
& + h(\{13\}, \{131\}) + h(\{13\}, \{132\}) + g(\{13\}) \notag\\
& + g(\{111\}) + g(\{112\})  + g(\{121\}) + g(\{122\})  + g(\{131\}) + g(\{132\}) .
\end{align}
Since the hopping is terminated at the boundary, the open boundary condition (OBC) is imposed naturally.

Now we shall see the behavior of local observables in the ground state under OBC.
Since the Hamiltonian \eqref{H_OBC} contains no interactions between particles, we can obtain the one-particle eigenstate from the following eigenvalue relation
\begin{align}
H \sum_{R} \psi_i(R) c^{\dag}_{R} \ket{0} = E_i \sum_{R} \psi_i(R) c^{\dag}_{R} \ket{0},
\label{eigen}
\end{align}
where $E_i$ is the $i$-th eigenvalue, and where $\psi_i(R)$ is the corresponding one-particle wave function.
We assume the increasing order for $E_i$.
By means of $\psi_i(R)$, the site occupation is expressed as
\begin{align}
n_R \equiv \braket{c^{\dag}_R c^{}_R} = \sum_{i(E_i \leq 0)} \psi_i^* (R) \psi_i (R)
\label{density}
\end{align}
under the grand canonical ensembles.
Similarly, the bond energy between neighboring sites $R$ and $R'$ can be obtained as
\begin{align}
\epsilon^{}_{R,R'} & \equiv -t \braket{c^{\dag}_{R} c^{}_{R'} + c^{\dag}_{R'} c_{R} } \notag \\
& = -t\sum_{i(E_i \leq 0)} \Bigl[ \psi_i^* (R) \psi_i (R') + \psi_i^* (R') \psi_i (R) \Bigr].
\label{bond_energy}
\end{align}

The nearest-neighbor hopping model on the tree has many degenerate energy levels due to the symmetry of the tree.
If and only if the chemical potential $\mu$ coincides with one of such energy levels, there exist a ground states degeneracy.
Throught this article, we avoid such a coincidence, and we choose $\mu$ so that the ground state is determined uniquely.
In this situation, the particle density $n_{R}$ in Eq.~\eqref{density} and the bond energy $\epsilon^{}_{R,R'}$ in Eq.~\eqref{bond_energy} are constant within each shell.
Hence, they are the functions of $s$, the shell index.
We thus use simpler forms
\begin{align}
n_s &\equiv n_{R_s},\\
\epsilon^{}_{s} & \equiv \epsilon^{}_{R_{s}, R_{s+1}},
\end{align}
where $R_s$ denotes $\{ r_0\: r_1 \ldots r_s \}$.

Since the total number of sites in the system increases exponentially with respect to $M$, it is not straight forward to solve the eigenvalue relation Eq.~\eqref{eigen} numerically when $M$ is relatively large.
This problem can be overcome by the block-diagonalization process\cite{Ogawa,Lepetit}.
All the blocks in the Hamiltonian matrix are symmetric tridiagonal matricies, whose size is $(M+1) \times (M+1)$ at most, so that one can easily diagonalize each.
Note that this block-diagonalization procedure can be applied also for those cases where Hamiltonian terms are position dependent, if $t$ and $\mu$ are the same for all the bonds and sites in the same shell.

\begin{figure}[htb]
\centering
\begin{minipage}{0.49\hsize}
\centering
\subfigure[]{ \includegraphics[width=\linewidth,clip]{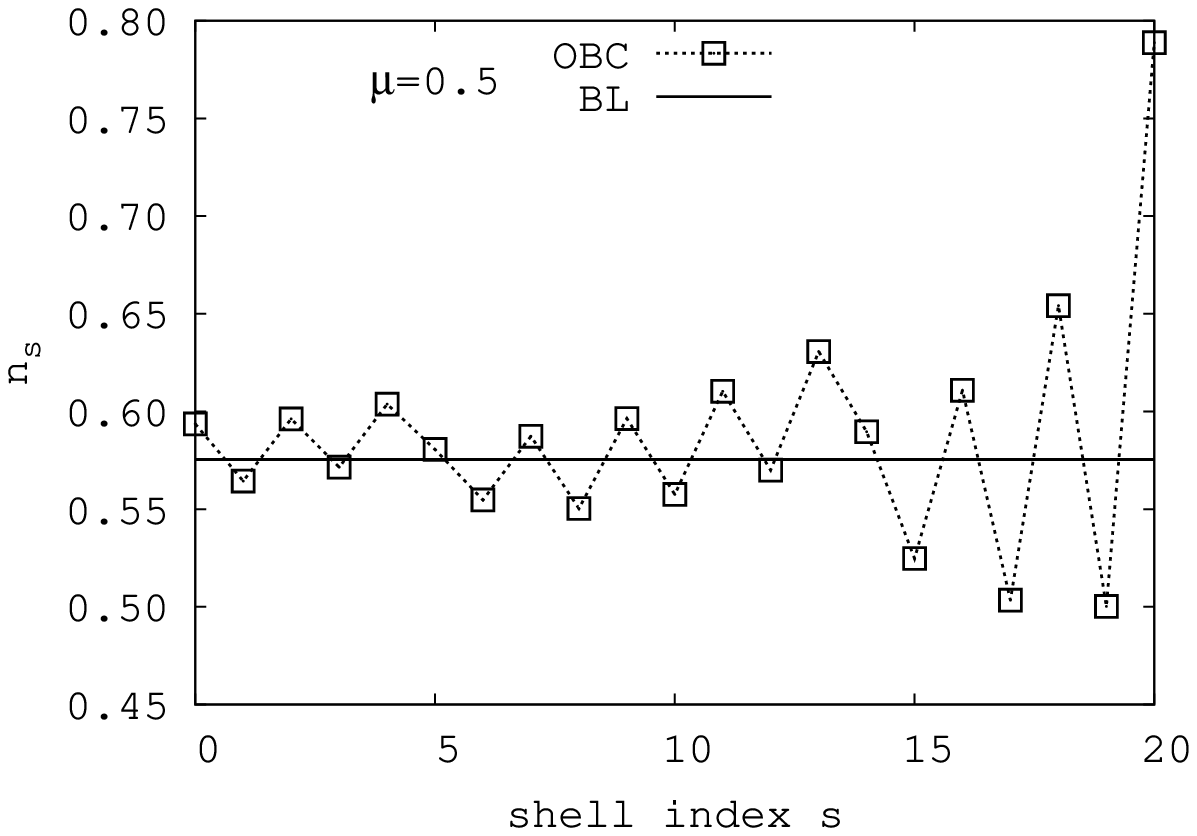} \label{ns_obc} }
\end{minipage}
\begin{minipage}{0.49\hsize}
\centering
\subfigure[]{ \includegraphics[width=\linewidth,clip]{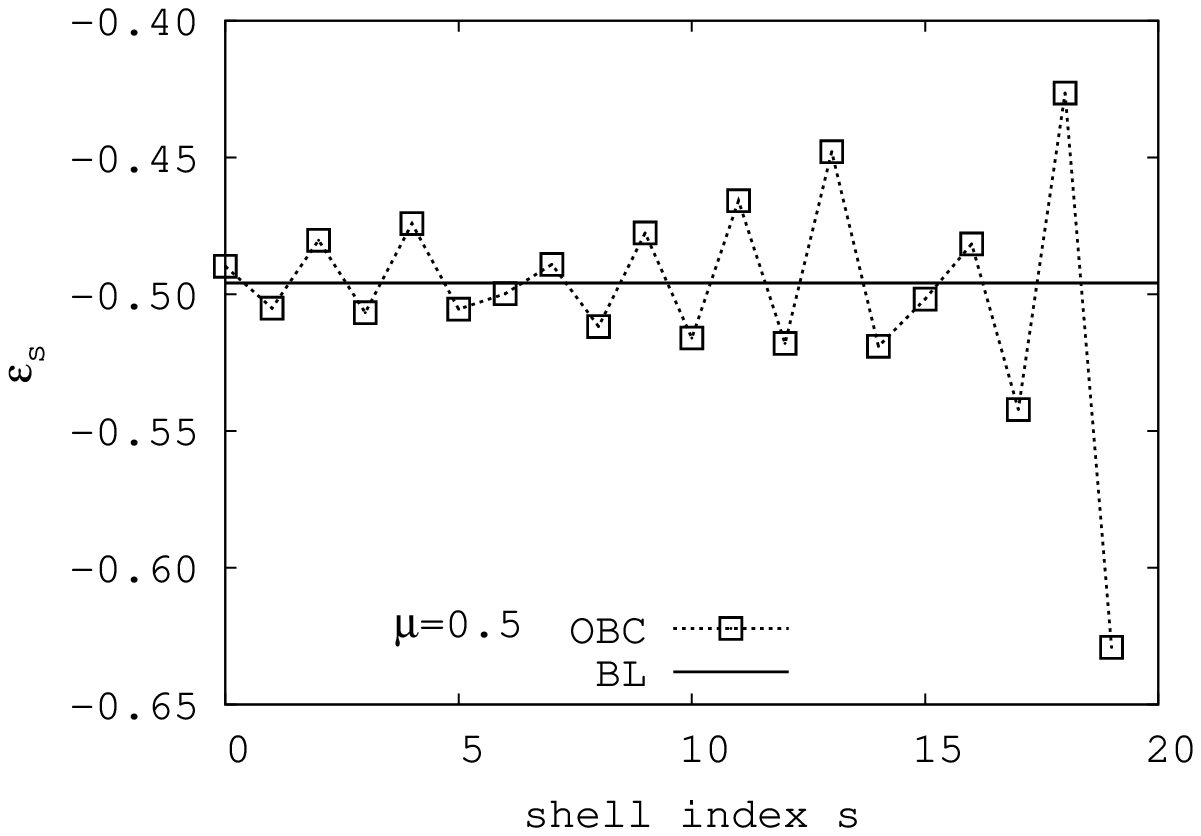} \label{es_obc} }
\end{minipage}
\caption{
The spatial dependence in
(a) the particle density $n_s$ and (b) the bond energy $\epsilon_s$,
on the tree with $z=3$ and $M=20$, when $\mu=0.5$.
We consider the open boundary condition.
The solid lines denote the corresponding quantities on the Bethe lattice.
}
\end{figure}
We shall compare the calculated density $n_s$ under OBC with that on the Bethe lattice\cite{Economou, Thorpe, Mahan, Eckstein, Kollar}, which is given by
\begin{align}
n (\mu) = \frac{1}{2\pi} \left[ z\theta_{\mu} - (z-2) \arctan \left( \frac{z}{z-2} \tan \theta_{\mu} \right) \right],
\label{exact-n}
\end{align}
where the relation $\mu = -2|t|\sqrt{z-1}\cos \theta_{\mu}$ is satisfied.
We also compare the bond energy $\epsilon^{}_{s}$ with that of the Bethe lattice
\begin{align}
\epsilon (\mu) = -\frac{|t|}{\pi} \left[ \sqrt{4(z-1)- (\mu /t)^2} - (z-2) \arctan \left( \frac{\sqrt{4(z-1)-(\mu /t)^2}}{z-2} \right) \right].
\label{exact-e}
\end{align}
In the following numerical calculations, we consider $t$ as the unit of energy.
Figure~\ref{ns_obc} shows $n_s$ when $z=3, M=20$ and $\mu=0.5$.
The solid line shows the corresponding bulk value in Eq.~\eqref{exact-n}.
We observe that $n_s$ oscillates around this bulk value and the oscillation amplitude is not negligible even at the center.
The similar fluctuation is observed in the bond energy as shown in Fig.~\ref{es_obc}.

\section{Smooth Boundary Condition} \label{sec:SBC}
\begin{figure}[hb]
\centering
\includegraphics[width=0.5\linewidth,clip]{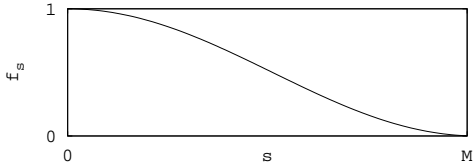}
\caption{The modulation function $f_s$ in Eq.~\eqref{ssd-function}.}
\label{fig-ssd}
\end{figure}

In this section, we consider the nearest-neighbor hopping model on the Cayley tree under SBC.
We start from the simplest case $z=2$, i.e., one-dimensional one.
Since there are only sites that are labeled such as $\{ 111\ldots 1 \}$ and $\{ 121\dots 1\}$ on the tree, the Hamiltonian \eqref{H_OBC} is reduced as
\begin{align}
H = \sum_{s=0}^{M-1} \sum_{r_1=1}^{2} h(\{ \underbrace{1\: r_1\: 1\ldots 1}_{s+1} \}, \{ \underbrace{1\: r_1\: 1 \ldots 1}_{s+2} \}) +g(\{1\}) + \sum_{s=1}^M \sum_{r_1=1}^{2} g(\{ \underbrace{1\: r_1\: 1\ldots 1}_{s+1} \}).
\end{align}
In the SBC scheme, local energy scale is smoothly modulated by a function $f_s$, where $s$ is the distance from the center of the system.
The modulation function $f_s$ is chosen to be smooth with respect to $s$ and vary from the maximum value near the center to zero at the edge.
Under such a setup, the Hamiltonian with SBC is written by
\begin{align}
H_{\mathrm{SBC}} = \sum_{s=0}^{M-1} f_{s+\frac{1}{2}} \sum_{r_1=1}^{2} h(\{ \underbrace{1\: r_1\: 1\ldots 1}_{s+1} \}, \{ \underbrace{1\: r_1\: 1 \ldots 1}_{s+2} \}) \notag\\
+ f_{0}\: g(\{1\}) + \sum_{s=1}^M f_{s} \sum_{r_1=1}^{2} g(\{ \underbrace{1\: r_1\: 1\ldots 1}_{s+1} \}).
\label{deformed_z2}
\end{align}
Now the generalization to those cases $z\geq 3$ is straight forward.
The genelized Hamiltonian can be rewritten as
\begin{align}
H_{\mathrm{SBC}} = \sum_{s=0}^{M-1} f_{s+\frac{1}{2}} \sum_{r_0\ldots r_{s+1}} h(\{ r_0\ldots r_s \}, \{ r_0\ldots r_{s+1} \}) +\sum_{s=0}^{M} f_s \sum_{r_0\ldots r_s} g(\{ r_0\ldots r_s \}).
\end{align}
As a choice of the smoothing function $f_s$, we choose 
\begin{align}
f_s = \sin^2 \left[ \frac{\pi}{2M+1} \left( s+M+\frac{1}{2} \right) \right],
\label{ssd-function}
\end{align}
where the functional form is shown in Fig.~\ref{fig-ssd}.
The deformation with this $f_s$ is called as sine-square deformation (SSD).

\begin{figure}[htb]
\centering
\begin{minipage}[b]{0.49\hsize}
\centering
\subfigure[]{ \includegraphics[width=\linewidth,clip]{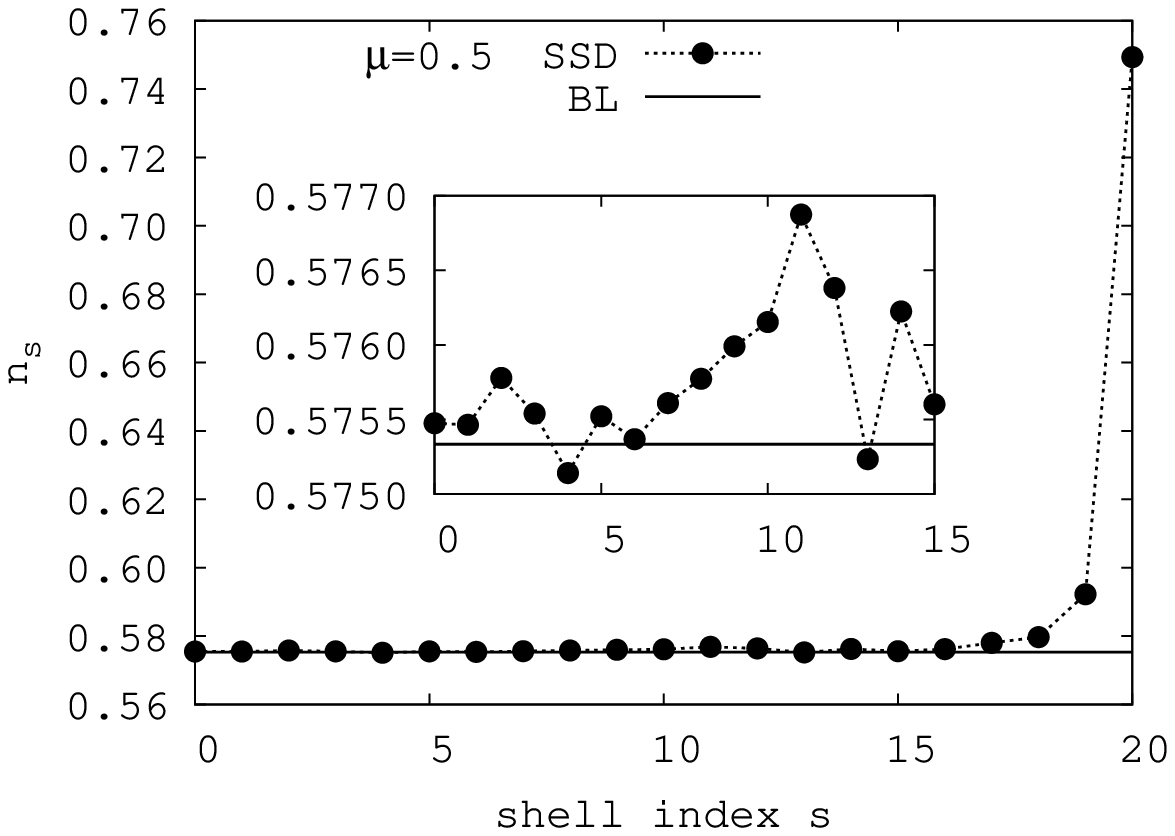} \label{ns_ssd} }
\end{minipage}
\begin{minipage}[b]{0.49\hsize}
\centering
\subfigure[]{ \includegraphics[width=\linewidth,clip]{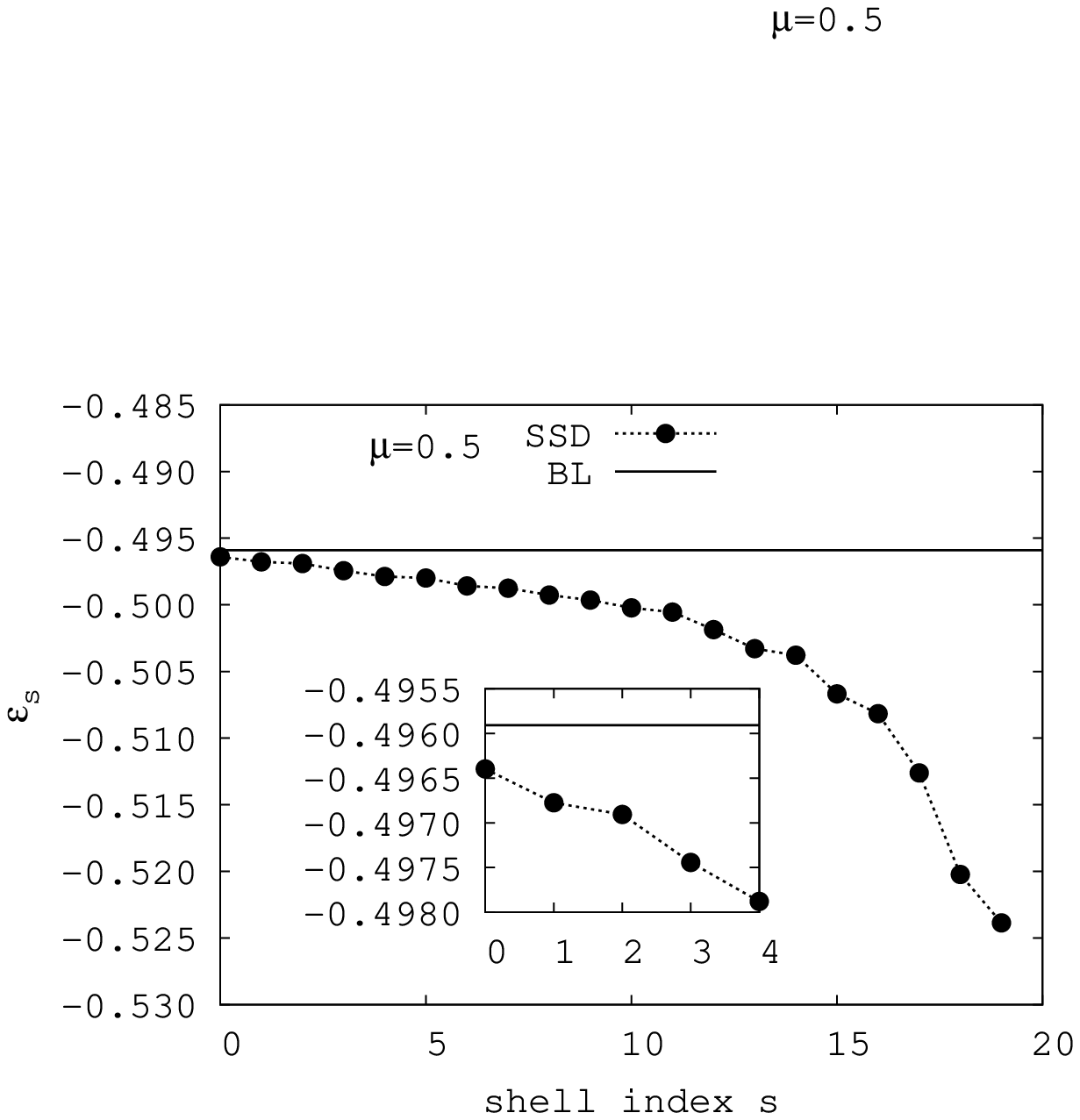} \label{es_ssd} }
\end{minipage}
\caption{
(a) The density of particles $n_s$ for the Cayley tree with $z=3$ and $M=20$ under SSD, obtained by Eq.~\eqref{ssd-function}.
We set the chemical potential as $\mu=0.5$.
(b) The density of particles $\epsilon_s$ under SSD. The parameter set is the same as (a).
The solid lines in both figures show the corresponding quantities on the Bethe lattice.
}
\end{figure}
Figure.~\ref{ns_ssd} shows the particle density $n_s$ under SSD with $z=3$, $M=20$, and $\mu=0.5$, where the parameter set is equivalent to those in Fig.~\ref{ns_obc}.
In the bulk region ($s\sim 1$), the oscillation is suppressed conspicuously, and the converged value well agrees with the bulk value on the Bethe lattice.
The diffrence lies within $O(10^{-4})$ near the center as shown in the inset.
SSD also reduces the oscillation in the bond energy $\epsilon_s$ as shown in Fig.~\ref{es_ssd}.
The calculated value approaches the bulk one toward the center.
The difference converge to the order of $10^{-4}$ as shown in the inset.

\begin{figure}[htb]
\begin{minipage}{0.49\hsize}
\centering
\subfigure[]{ \includegraphics[width=\linewidth,clip]{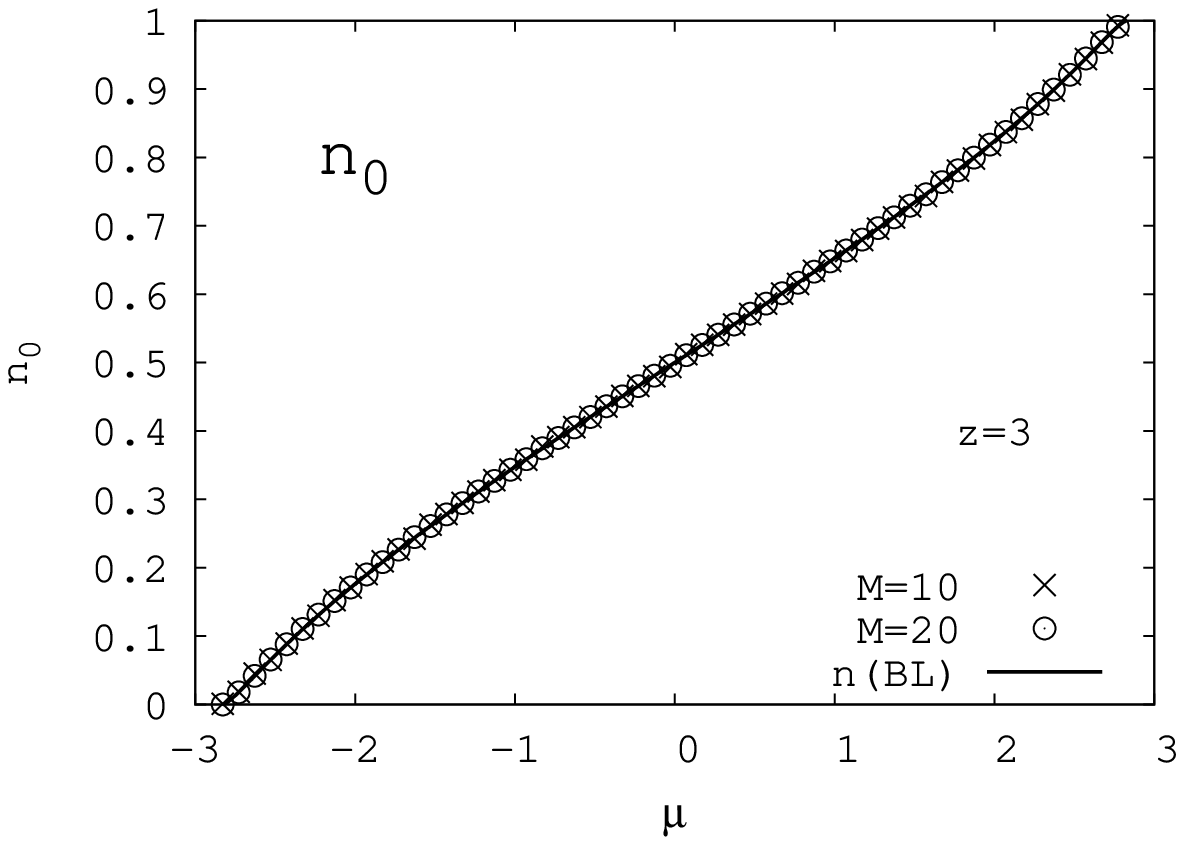} \label{n0} }
\end{minipage}
\begin{minipage}{0.49\hsize}
\centering
\subfigure[]{ \includegraphics[width=\linewidth,clip]{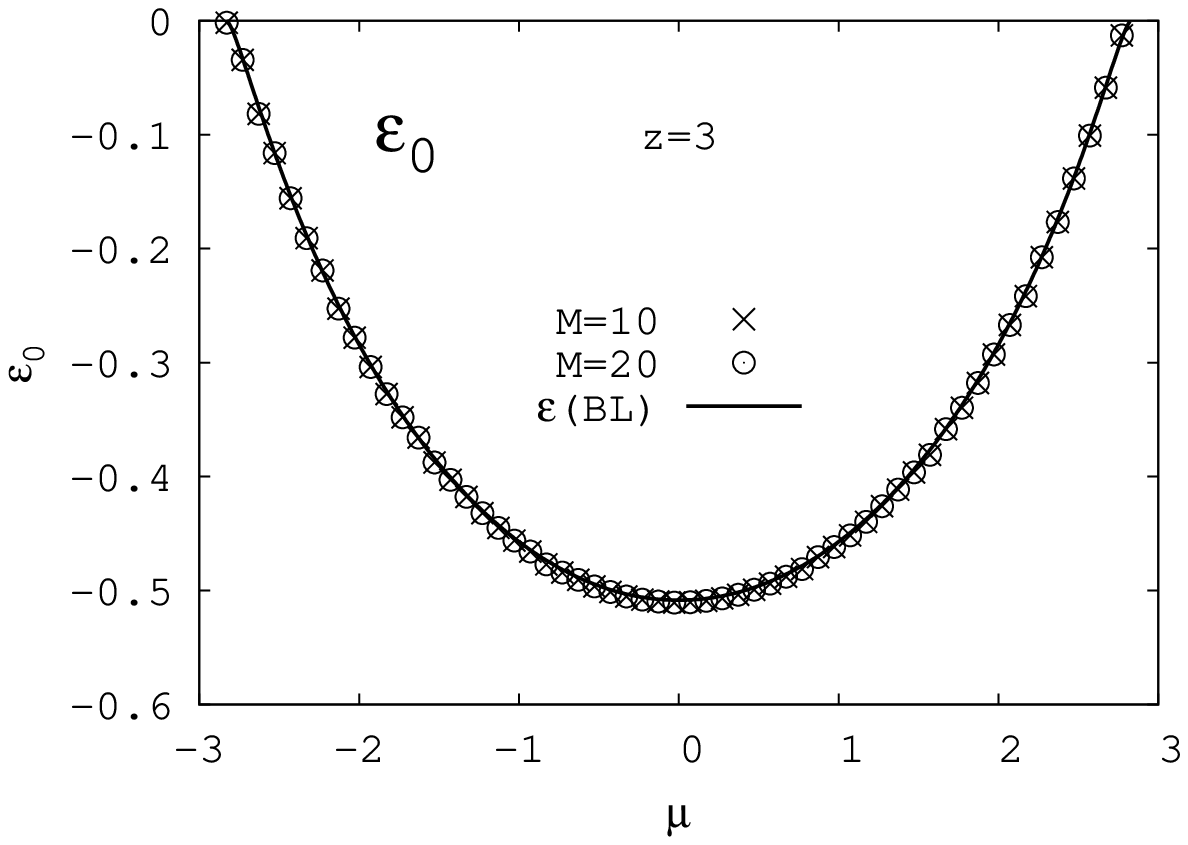} \label{e0} }
\end{minipage}
\caption{
The $\mu$ dependence in (a) the particle density at the center $n_0$, and (b) the bond energy at the center $\epsilon_0$.
The solid line in (a) and (b) show the bulk value $n(\mu)$ and $\epsilon (\mu)$ on the Bethe lattice, respectively, in Eq.~\eqref{exact-n} and \eqref{exact-e}.
}
\end{figure}
Now we estimate the bulk density of the system by the ground canonical analysis applied to the calculated data.
As seen in Fig.~\ref{ns_ssd}, the oscillation around the central site of the system is small.
Thus we may consider $n_0$ as the candidate for the bulk density.
We compare the obtained density $n_0$ with that on the Bethe lattice $n(\mu)$, which is given in Eq.~\eqref{exact-n}.
Figure~\ref{n0} shows the $\mu$ dependence in $n_0$.
The curves of $n_0$, calculated for $M=10$ and 20, well coincides with $n(\mu)$, where the typical discrepancy is at most of the order of $10^{-3}$.
We also compare $\epsilon_0$ with the bond energy on the Bethe lattice $\epsilon (\mu)$ in Eq.~\eqref{exact-e}, in the same manner.
The coincidence is observed as shown in Fig.~\ref{e0}.
The difference between $\epsilon_0$ and $\epsilon (\mu)$ lies within the order of $10^{-3}$.

\section{Conclusion} \label{sec:last}
We have applied SSD to the nearest-neighbor hopping model on the Cayley tree.
As an effect of the boundary condition, the particle density become relatively uniform in the bulk region, deep inside the tree.
The $\mu$ dependence of the particle density at the center of the system coincide with that on the Bethe lattice even for a small $M$, the number of shells.
We also observe such coincidence in the bond energy.
These fact suggest that SSD enables us to extract bulk quantities of various local observables for the Cayley tree in an efficient way.
A future problem is to examine other deformations such as sinusoidal\cite{ssd-GDLN,Hikihara}, exponential\cite{exp1,exp2} and hyperbolic\cite{hyp1,hyp2,hyp3} deformation.

\section*{Acknowledgement}
The author would like to thank T. Nishino for valuable comments.


\end{document}